\documentclass{article}
\usepackage[a4paper, top=35truemm, bottom=30truemm, left=24truemm, right=24truemm]{geometry}
\usepackage{graphicx}
\usepackage{xcolor}
\usepackage{physics}
\usepackage{amssymb}
\usepackage{mathrsfs}
\usepackage{tensor}
\usepackage{mathtools}

\usepackage{enumitem}
\usepackage{authblk}
\usepackage{pdfpages}
\usepackage{cite}

\usepackage{hyperref}
\hypersetup{colorlinks=true, allcolors=blue}

\DeclareMathOperator\Arccos{Arccos} 
 
\DeclareMathOperator\Log{Log} 

\newcommand{\p}{{$p_\circ\,$}}
\newcommand{\rp}{{r_{(+)}}}
\newcommand{\rn}{{r_{(-)}}}
\newcommand{\D}{{\cal D}}

\begin{document}

\title{Multi-Sheet Wormholes in the Gravitational Soliton Formalism}
\author[1]{Yusuke Makita}
\author[2,3]{Keisuke Izumi}
\author[2]{Daisuke Yoshida}
\author[1]{Keiya Uemichi}
\affil[1]{Division of Science, Graduate School of Science, Nagoya University, Nagoya 464-8602, Japan}
\affil[2]{Department of Mathematics, Nagoya University, Nagoya 464-8602, Japan}
\affil[3]{Kobayashi-Maskawa Institute, Nagoya University, Nagoya 464-8602, Japan}
\date{}
\maketitle
\vspace{-10mm}
\begin{tabbing}
    \hspace{10mm} E-mail: \ \= \href{mailto:makita.yusuke.c2@s.mail.nagoya-u.ac.jp}{makita.yusuke.c2@s.mail.nagoya-u.ac.jp}, \\ \>
    \href{mailto:izumi@math.nagoya-u.ac.jp}{izumi@math.nagoya-u.ac.jp}, \\ \>
    \href{mailto:dyoshida@math.nagoya-u.ac.jp}{dyoshida@math.nagoya-u.ac.jp}, \\ \>
    \href{mailto:uemichi.keiya.j4@s.mail.nagoya-u.ac.jp}{uemichi.keiya.j4@s.mail.nagoya-u.ac.jp}
\end{tabbing}

\vspace{5mm}
\begin{abstract}
We analytically construct static regular solutions describing wormholes that connect multiple asymptotic regions, supported by a phantom scalar field.  
The solutions are static and axially symmetric, and are constructed using the gravitational soliton formalism, in which the equations of motion reduce to the Laplace equations on a two-dimensional sheet. However, the presence of multiple asymptotic regions necessitates the introduction of multiple such sheets. These sheets are appropriately cut and glued together to form a globally regular geometry. This gluing procedure represents the principal distinction from conventional Weyl-type solitonic solutions and is a characteristic feature of the wormhole geometries studied in this paper.
\end{abstract}

\section{Introduction}
In the framework of general relativity, solutions with non-trivial geometry, such as {\it wormholes} connecting distinct asymptotic regions, have been widely studied.
The earliest example was discovered by Einstein and Rosen, who pointed out that a bridge-like structure exists in the Schwarzschild spacetime~\cite{EinsteinRosen}. This marked the beginning of wormhole research; however, the structure is not \textit{traversable}.
In the 1980s, Morris, Thorne, and Yurtsever investigated the possibility of traversable wormholes~\cite{MorrisThorne, MorrisThorne2}.
Since then, various wormhole solutions have been constructed, including those supported by a phantom field~\cite{Ellis1973, Bronnikov,Picon2002,Carvente:2019gkd,Chew:2019lsa,Nozawa:2023aep}, in modified gravity~\cite{Maeda:2008nz, Myrzakulov:2015kda}, in higher dimensions~\cite{Torii:2013xba, Baruah:2019cfg, Martinez:2020hjm, Nozawa:2020wet}, those supported by Casimir energy~\cite{DelAguila:2015isj,Maldacena:2018gjk}, and within the braneworld context~\cite{Ida:2001qw, Tomikawa:2014wxa}.
Efforts have also been made to define wormholes in dynamical spacetimes~\cite{Hochberg:1997wp, Hayward:1998pp,
Hayward:2002pm,
Koyama:2004uh,
Hayward:2009yw,
Maeda:2008bh,Tomikawa:2015swa}, and wormhole solutions have been constructed within cosmological frameworks~\cite{Maeda:2009tk}.
A detailed history of these developments can be found, for instance, in the textbook by Visser~\cite{MattVisser}.

The simplest way to support the structure of a traversable wormhole is to introduce a \textit{phantom} scalar field, which possesses negative kinetic energy.
The first such solution was independently presented by Ellis and Bronnikov in 1973~\cite{Ellis1973, Bronnikov}; it describes a static, spherically symmetric wormhole with vanishing  Arnowitt--Deser--Misner (ADM) mass. A generalization that includes ADM mass was studied in Ref.~\cite{Picon2002}.
This paper presents a systematic method for constructing static, axially symmetric solutions supported by a phantom scalar field, and introduces new examples featuring multiple asymptotic regions.

Our idea behind constructing new analytic wormhole solutions is to start with an axisymmetric ansatz.
The earliest and most well-known systematic construction of axisymmetric solutions was presented by Weyl for static spacetimes, 
leading to the \textit{Weyl class} of solutions~\cite{Weyl1917,Emparan:2001wk, Griffiths_Podolsky_2009}.
In this class, most vacuum solutions exhibit naked singularities, although they can still be interpreted as exterior solutions of certain sources~\cite{Saito:2024hzc}.
For stationary spacetimes, the Ernst formalism~\cite{Ernst1968,Ernst1968b} and the gravitational soliton formalism~\cite{Belinski1978, Belinski1979}  provide useful frameworks for constructing solutions, 
and they offer systematic methods for obtaining the Kerr solution.
Moreover, these methods have also produced non-trivial five-dimensional black hole solutions~\cite{Emparan:2001wk,Emparan:2001wn,Elvang:2004rt,Harmark:2004rm,Tomizawa:2005wv,Mishima:2005id,Figueras:2005zp,Tomizawa:2006vp,Iguchi:2006rd,Pomeransky:2006bd,Emparan:2006mm,Elvang:2007hs,Elvang:2007rd,Tomizawa:2007mz,Iguchi:2007xs,Evslin:2007fv,Izumi:2007qx,Tomizawa:2008qr,Evslin:2008gx,Chen:2008fa,Iguchi2011,Iguchi:2011qi,Rocha:2011vv,Chen:2012kd,Chen:2012zb,Feldman:2012vd,Rocha:2012vs,Tomizawa:2016kjh,Tomizawa:2019acu,Chew:2019lsa,Tomizawa:2022qyd,Suzuki:2023nqf,Suzuki:2024coe,Suzuki:2024phv}, 
which have stimulated increased research activity in 
higher dimensional black hole physics.
In this paper, we apply one of these powerful methods, the gravitational soliton formalism, to construct wormhole solutions, 
while the Ernst formalism is discussed in the work of Volkov~\cite{Volkov2021}.

\begin{figure}[t]
    \centering
    \includegraphics{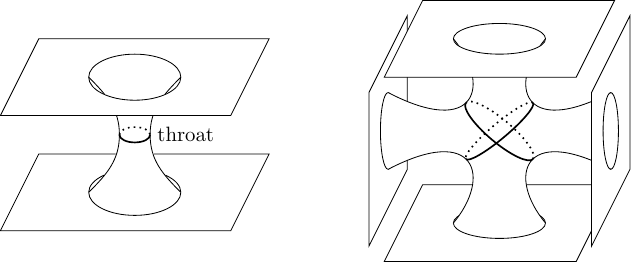}
    \caption{The left figure depicts a standard wormhole spacetime with two asymptotic regions, while the right figure illustrates a \textit{multi-sheet} wormhole with four asymptotic regions. The construction of such multi-sheet wormholes is one of the main results of this paper.}
    \label{fig:wormhole0}
\end{figure}

Following the gravitational soliton formalism presented by Belinski and Zakharov~\cite{Belinski1978, Belinski1979}, we decompose stationary and axially symmetric spacetimes into two parts: the directions $t$ and $\phi$, corresponding to stationarity and axial symmetry, respectively, and the remaining spatial directions $\rho$ and $z$. 
In the static case, the field equations for the logarithms of the metric functions $\ln\abs{g_{tt}}$, $\ln g_{\phi\phi}$, and the scalar field $\Phi$ reduce to the Laplace equation in a three-dimensional Euclidean space with axial symmetry. 
Since $\ln\abs{g_{tt}}$, $\ln g_{\phi\phi}$, and $\Phi$ are solutions of the Laplace equation, they can be expressed as superpositions of functions called \textit{solitons} $\mu_k$.
Reformulation of the spherically symmetric wormhole solutions~\cite{Ellis1973,Bronnikov, Picon2002}
within the gravitational soliton formalism
yields an explicit expression for the solitons that describe wormholes. These solitons, in turn, enable the construction of 
wormhole geometries that connect multiple asymptotic regions (see Fig.~\ref{fig:wormhole0}).

The major difference between the conventional soliton formalisms and the approach taken in this paper is that, for wormholes, it is necessary to introduce multiple $(\rho, z)$ sheets. Hence, we refer to these geometries as \textit{multi-sheet wormhole}. 
In the soliton formalism, the equations for certain metric components and scalar fields reduce to the Laplace equations in a three-dimensional Euclidean space, constructed from a two-dimensional $(\rho, z)$ sheet and an auxiliary axially symmetric direction. 
In the conventional soliton formalisms, which were used for constructing higher-dimensional black hole solutions, the analysis is confined to a single $(\rho, z)$ sheet.
In contrast, wormholes possess multiple sheets.
These sheets contain cuts and are connected across them.
A careful analysis of the regularity of these structures is necessary, and the details are presented in this paper.

This paper is organized as follows. In Section~\ref{basics}, we review the gravitational soliton formalism and present the equations for the static case.  
In Section~\ref{twosheet}, we reformulate known, spherically symmetric two-sheet wormhole solutions with nonzero mass~\cite{Picon2002} within the gravitational soliton formalism and figure out the structure of the solitons for describing wormhole geometries.
We also present the regularity conditions required to connect the sheets.
In Section~\ref{multisheet}, we construct multi-sheet wormhole solutions.
Finally, in Section~\ref{summary}, we provide a brief summary of this paper.  
In Appendix~\ref{solitons}, we show a relation between the solitons of black holes and those of wormholes.

\section{Solitonic formulation for static and axially symmetric solutions} \label{basics}
In this section, we review a method for constructing static and axially symmetric solutions of the Einstein equations with a massless scalar field,
which is also applicable to the theory with a massless phantom field.
We will apply this method in the following sections.
An extension to the stationary case is straightforward by following the method of Belinski and Zakharov~\cite{Belinski1978, Belinski1979} (see also a review~\cite{Iguchi:2011qi} and a recent work~\cite{Vigano:2022hrg}),
although we do not present it in this paper.

\subsection{Field equations} \label{fieldeqs}
Our main objective is to construct an exact solution for a four-dimensional, static, and axially symmetric spacetime. 
The line element of stationary and axially symmetric spacetime can be written in
\begin{equation}
    \dd{s}^2 \coloneqq g^{(4)}_{\mu\nu}\dd{x}^{\mu}\dd{x}^{\nu} \coloneqq g_{ab}\qty(y^{i})\dd{x}^{a}\dd{x}^{b} + \dd{l}^2, \label{eq:metric2} \\
\end{equation}
with
\begin{equation}
    \dd{l}^2 \coloneqq \gamma_{ij} (y^i) \dd{y}^{i} \dd{y}^{j}, \label{eq:dl}
\end{equation}
where $g^{(4)}_{\mu\nu}$ is the four-dimensional full metric, and we denote $x^{a}=\qty(t, \phi)$ as coordinates adapted to the static and axial symmetries, and $y^{i}$ as the rest two-dimensional spatial components.
We assume that the symmetry along $t$ is timelike, $g_{tt} < 0$, and that along $\phi$ is spacelike, $g_{\phi\phi}> 0$.
The metric components $g_{ab}$ and $\gamma_{ij}$ depend only on $y^{i}$, because of the symmetries. 

Let us consider the Einstein--Hilbert action with a massless scalar field $\Phi$,
\begin{equation}
    S\qty[g^{(4)}_{\mu\nu}, \Phi] \coloneqq \frac{1}{2\kappa}\int\qty[R^{(4)} - 2g^{(4)\mu\nu}\qty(\partial_{\mu}\Phi)\qty(\partial_{\nu}\Phi)]\sqrt{-g^{(4)}}\dd[4]{x}. \label{eq:action1}
\end{equation}
Note that the overall sign of the scalar field action is chosen so that the scalar field has positive kinetic energy when $\Phi$ is real; a phantom scalar field is represented by taking $\Phi$ to be purely imaginary.
This action gives the energy momentum tensor $T_{\mu\nu}$ for the scalar field,
\begin{equation}
    \kappa T_{\mu\nu} = 2\qty(\partial_{\mu}\Phi)\qty(\partial_{\nu}\Phi) - g^{(4)\alpha\beta}\qty(\partial_{\alpha}\Phi)\qty(\partial_{\beta}\Phi)g^{(4)}_{\mu\nu}.
\end{equation}
The Einstein field equation can be simplified as
\begin{equation}
    R^{(4)}_{\mu\nu} = 2\qty(\partial_{\mu}\Phi)\qty(\partial_{\nu}\Phi). \label{eq:einstein1}
\end{equation}

By the metric ansatz \eqref{eq:metric2}, with imposing the same symmetry for the scalar field $\Phi = \Phi(y^{i})$, 
the components of the Ricci tensor along the symmetry directions can be written as
\begin{equation}
    R^{(4)a}{}_b = -\frac{1}{2\sqrt{-g}}\,\D_{i}\qty(\sqrt{-g}g^{ac}\D^i g_{cb}), \label{eq:ricci1}
\end{equation}
where $\D_i$ denotes the covariant derivative on the two-dimensional space spanned by $y^i$. 
Meanwhile, the corresponding components of the right-hand side of Eq.~\eqref{eq:einstein1} vanish.
This implies
\begin{equation}
    \D^{i}\qty(\sqrt{-g}g^{ac}\D_{i}g_{cb}) = 0. \label{eq:ricci2}
\end{equation}
In particular, taking the trace of Eq.~\eqref{eq:ricci2} yields the two-dimensional Laplace equation:
\begin{equation}
    \D^{i}\D_{i}\sqrt{-g} = 0.
    \label{eq:gharm}
\end{equation}
Hence, $\sqrt{-g}$ is a harmonic function on the surface spanned by $y^{i}$, and one can therefore choose it as one of the coordinates $y^{i}$, namely $\rho$, that is,
\begin{equation}
    \sqrt{-g} = \rho. \label{eq:gauge1}
\end{equation}
We choose the other coordinate as the harmonic conjugate $z$ \footnote{The harmonic conjugate of $\rho$ is defined by $\partial_{i} z = \epsilon_{i}{}^{j} \partial_{j} \rho$ using the Levi-Civita tensor $\epsilon_{ij}$ associated with $\dd{l}^2$. We can choose this coordinate at least locally.} of $\rho$, 
and then the two-dimensional space spanned by $y^i$ can be expressed in a conformal flat metric,
\begin{equation}
    \dd{l}^2 = F(\rho,z) \left[\dd \rho^2 + \dd z^2 \right]. \label{eq:dl^2}
\end{equation}
Hereinafter in this section, raising and lowering spatial indices $i, j$ are performed by the two dimensional {\it flat} metric $\delta_{ij}$ spanned by $\rho$ and $z$.
In this coordinate system, Eq.~\eqref{eq:ricci2} becomes
\begin{equation}
    \partial^{i}\qty\big[\rho\,\partial_{i}\left(\ln g\right)_{ab}] = 0, \label{eq:laplace1}
\end{equation}
which is nothing but the axisymmetric form of the Laplace equation in three-dimensional Euclidean space, written in cylindrical coordinates.
For convenience, we introduce two auxiliary fields:
\begin{equation}
    \begin{split}
        U\indices{^a_b} &\coloneqq \rho\, g^{ac}\partial_{\rho}g_{cb}, \\
        V\indices{^a_b} &\coloneqq \rho\, g^{ac}\partial_{z}g_{cb}.
    \end{split}
    \label{eq:defUV}
\end{equation}
With these fields, Eq.~\eqref{eq:laplace1} can be rewritten as
\begin{equation}
    \partial_{\rho}U\indices{^a_b} + \partial_{z}V\indices{^a_b} = 0. \label{eq:integra1}
\end{equation}

The action~\eqref{eq:action1} yields  the massless Klein-Gordon equation for $\Phi$,
\begin{equation}
    \nabla_{\mu}\nabla^{\mu}\Phi = 0. \label{eq:kg1}
\end{equation}
The scalar field is supposed to have the same symmetries as the spacetime,
that is, $\Phi = \Phi(\rho,z)$,
thereby reducing the Klein-Gordon equation to the Laplace equation:
\begin{equation}
\partial^{i}\left(\rho\partial_{i}\Phi\right) = 0. \label{eq:scalar1}
\end{equation}
Note that this equation has the same structure as Eq.~\eqref{eq:laplace1}.

The conformal factor $F(\rho, z)$ for the $(\rho, z)$ surface is determined by the remaining components in the Einstein equation~\eqref{eq:einstein1},
which lead to
\begin{equation}
    \left\{ 
    \begin{alignedat}{3}
        \partial_{\rho}\ln F &= -\frac{1}{\rho} +{} & & \frac{1}{4\rho}\left(U\indices{^a_b}U\indices{^b_a} - V\indices{^a_b}V\indices{^b_a}\right) & & + 2\rho\left[\left(\partial_{\rho}\Phi\right)^2 - \left(\partial_{z}\Phi\right)^2\right], \\
        \partial_{z}\ln F &= & & \frac{1}{2\rho}U\indices{^a_b}V\indices{^b_a} & & + 4\rho\left(\partial_{\rho}\Phi\right)\left(\partial_{z}\Phi\right).
    \end{alignedat}
    \right. \label{eq:eqforF}
\end{equation}
Equation~\eqref{eq:integra1} guarantees the integrability condition for $\ln F$.

To summarize, under the coordinate choice~\eqref{eq:gauge1}, the equations of motion reduce to the Laplace equations for $\qty(\ln g)_{ab}$ and $\Phi$, namely Eqs.~\eqref{eq:laplace1} and \eqref{eq:scalar1}, together with the integrable equations for $F$ given in Eq.~\eqref{eq:eqforF}.

\subsection{Solitonic formulation of static solutions}
\begin{figure}[t]
    \centering
    \begin{minipage}[h]{0.45\linewidth}
        \centering
        \includegraphics{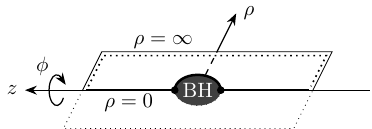}
        \caption{Cylindrical coordinates embedded in the black hole spacetime. The coordinate $\rho$ vanishes not only on the axis but also on the horizon. The $(\rho, z)$ coordinates only cover the static region; the outside of the horizon.}
        \label{fig:blackhole1}
    \end{minipage}%
    \begin{minipage}[h]{0.50\linewidth}
        \centering
        \includegraphics{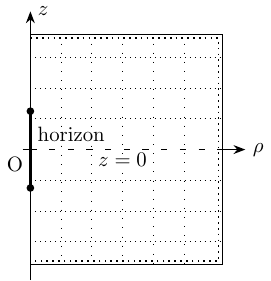}
        \caption{The black hole spacetime drawn in $\qty(\rho, z)$ coordinates. The horizon is expressed as a segment at $\rho = 0$.}
        \label{fig:blackhole2}
    \end{minipage}
\end{figure}

We now restrict our discussion to the static case.
In this situation, the metric components $g_{ab}$ along the symmetry directions take a diagonal form,
\begin{equation}
 g_{ab}(y^{i})\dd{x}^{a} \dd{x}^{b} = g_{tt}(\rho,z) \dd{t}^2 + g_{\phi\phi}(\rho,z) \dd{\phi}^2
=  - |g_{tt}(\rho,z)| \dd{t}^2 + \frac{\rho^2}{|g_{tt}(\rho,z)|} \dd{\phi}^2,
\end{equation}
which thus corresponds to the four-dimensional metric,
\begin{equation}
 \dd{s}^2 = - |g_{tt}(\rho,z)| \dd{t}^2 + \frac{\rho^2}{|g_{tt}(\rho,z)|} \dd{\phi}^2  + F(\rho,z) \qty[\dd{\rho}^2 + \dd{z}^2]. \label{eq:ds2static}
\end{equation}
Here, we use the coordinate choice given in Eq.~\eqref{eq:gauge1},
\begin{equation}
    \rho^2 = -\det g_{ab} = -g_{tt}g_{\phi\phi}. \label{eq:gauge2}
\end{equation}
The condition~\eqref{eq:gauge2} implies that the metric components $g_{ab}$ become degenerate in the limit $\rho \rightarrow 0$.
By requiring the finiteness of the curvature invariants,
there are two possible cases in which no singularity appears at $\rho = 0$,
\begin{equation}
    \left\{
    \begin{alignedat}{1}
        g_{tt} &\propto -\rho^2 \\
        g_{\phi\phi} &\not= 0
    \end{alignedat} 
    \right.
    \qquad \text{and} \qquad
    \left\{
    \begin{alignedat}{1}
        g_{tt} &\not= 0 \\
        g_{\phi\phi} &\propto \rho^2
    \end{alignedat}.
    \right.
\end{equation}
The former corresponds to the presence of a horizon, while the latter represents the axis of the rotational symmetry.
Any other behavior results in a singularity\cite{Harmark:2004rm}.
The three-dimensional configuration is illustrated in Fig.~\ref{fig:blackhole1}, 
and the location of the horizon and the axis in the $(\rho,z)$ coordinates are depicted in Fig.~\ref{fig:blackhole2}.
Note that the $(\rho, z)$ coordinates only cover the static region---that is, the exterior of the black hole.

From Eq.~\eqref{eq:laplace1},
the $(tt)$ and $(\phi\phi)$ components of metric are required to satisfy
\begin{align}
    \partial^{i}\left(\rho \, \partial_{i}\ln |g_{tt}|\right) &= 0, 
    \label{eq:gttlap}\\
    \partial^{i}\left(\rho \, \partial_{i}\ln g_{\phi\phi}\right) &= 0,
    \label{eq:gpplap}
\end{align}
while the $(t\phi)$ component of Eq.~\eqref{eq:laplace1} is trivially satisfied. 
These are three-dimensional Laplace equations for axially symmetric scalar fields, as in Eq.~\eqref{eq:scalar1}.
Therefore, the functions $g_{tt}$, $g_{\phi\phi}$, and $\Phi$ can be expressed as "superpositions" of solutions $\mu_k$ satisfying
\begin{equation}
    \partial^{i}\left(\rho \, \partial_{i} \ln \mu_k \right) = 0, \label{eq:soliton1}
\end{equation}
where $k$ labels the individual solutions rather than spacetime coordinates, namely,
\begin{equation}
    \begin{split}
        \ln |g_{tt}| &= a_1 \ln \mu_1 + a_2 \ln \mu_2 + \cdots, \\
        \ln g_{\phi\phi} &= a_3 \ln \mu_3 + a_4 \ln \mu_4 + \cdots, \\
        \Phi &= a_5 \ln \mu_5 + a_6 \ln \mu_6 + \cdots.
    \end{split}
    \label{eq:soliton2}
\end{equation}
We refer to the solutions $\mu_k$ as \textit{solitons} in the context of solution superposition.
Note that the metric components $g_{tt}$ and $g_{\phi\phi}$ are not independent, since the coordinate condition has been fixed in Eq.~\eqref{eq:gauge1}.

Once the solutions $|g_{tt}|, g_{\phi\phi}$ and $\Phi$ are obtained, the remaining metric component $F$ can be determined by integrating Eqs.~\eqref{eq:eqforF}, which now can be expressed as
\begin{equation}
    \left\{ 
    \begin{alignedat}{3}
        \partial_{\rho}\ln F &= -\partial_{\rho}\ln\abs{g_{tt}} & &+ \frac{\rho}{2} \qty[\qty(\partial_{\rho}\ln\abs{g_{tt}})^2 - \qty(\partial_{z}\ln\abs{g_{tt}})^2] & &+ 2\rho\qty[\qty(\partial_{\rho}\Phi)^2 - \qty(\partial_{z}\Phi)^2], \\
        \partial_{z}\ln F &= -\partial_{z} \ln\abs{g_{tt}} & &+ \rho\qty\Big[\qty(\partial_{\rho}\ln\abs{g_{tt}})\qty(\partial_{z}\ln\abs{g_{tt}})] & &+ 4\rho\qty\Big[\qty(\partial_{\rho}\Phi)\qty(\partial_{z}\Phi)].
    \end{alignedat}
    \right. \label{eq:eqforFstatic}
\end{equation}

\section{Two-sheet wormhole solutions} \label{twosheet}

In this section, we reformulate a class of static, spherically symmetric wormhole solutions connecting two asymptotic regions originally obtained in Ref.~\cite{Picon2002}, using the solitonic formulation presented in the previous section.
Although the original solutions are expressed in the spherically symmetric coordinates, they can be rewritten in the axisymmetric form~\eqref{eq:ds2static}, allowing us to uncover the solitonic structure of the wormhole spacetime.
This reformulation naturally leads to a generalization to wormhole solutions with multiple asymptotic regions, which will be discussed in the next section.

In Sec.~\ref{TSWtop}, we describe the structure of the $(\rho, z)$ sheet.
Because the solution has two spatial infinities, two $(\rho, z)$ sheets must be introduced.
The solitonic representation of the wormhole solution is extracted in Sec.~\ref{soliton}.
Since the two $(\rho, z)$ sheets are eventually connected,
it is necessary to verify the regularity of this connection, in addition to the regularity conditions along the axis,
which are discussed in the standard solitonic solutions of Refs.~\cite{Belinski1978,Belinski1979}.
This analysis of the regularity conditions is carried out in Sec.~\ref{regularity}.

\subsection{Wormhole geometries}\label{TSWtop}

\begin{figure}[t]
    \centering
    \begin{minipage}[h]{0.42\linewidth}
        \centering
        \includegraphics{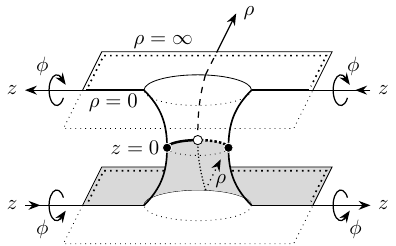}
        \caption{Cylindrical coordinates embedded in a wormhole spacetime.
        Each sheet (upper and shaded lower) corresponds to the asymptotically flat regions, and is individually spanned by $(\rho, z)$ coordinates.}
        \label{fig:wormhole1}
    \end{minipage}
    \hspace{0.03\linewidth}
    \begin{minipage}[h]{0.5\linewidth}
        \centering
        \includegraphics{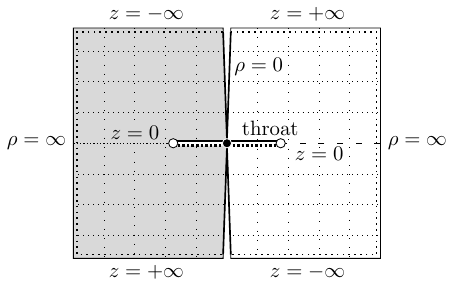}
        \caption{The wormhole spacetime drawn with two $\qty(\rho, z)$ sheets. 
        The right sheet corresponds to the upper sheet in Fig.~\ref{fig:wormhole1}, while the shaded left represents the shaded lower.
        Each sheet in Fig.~\ref{fig:wormhole1} is deformed around the point $\circ$ and is glued together along the solid and dotted curves, resulting in each sheet in this figure.}
        \label{fig:wormhole3}
    \end{minipage}
\end{figure}

A wormhole spacetime connecting two distinct spatial infinities is illustrated in Fig.~\ref{fig:wormhole1}.  
As in the black hole case shown in Fig.~\ref{fig:blackhole1}, the axial direction $\phi$ is suppressed for clarity.  
The wormhole contains two asymptotic regions.
They are divided by a surface, which we call \textit{throat}.
From the structure of the surface shown in Fig.~\ref{fig:wormhole1}, one can cut the surface along a horizontal curve connecting the points $\bullet$, $\circ$, and $\bullet$ in the center of the figure. This operation divide the surface into two sheets, corresponding to the two asymptotic regions.

Let us focus on the upper sheet of Fig.~\ref{fig:wormhole1}.  
We divide the cutting curve into two parts: the solid (heavy) curve between $\bullet$ and $\circ$, and the dotted curve between $\circ$ and the other $\bullet$.  
By continuously deforming the surface and bringing these two curves closer together, we eventually obtain the right-hand sheet shown in Fig.~\ref{fig:wormhole3}.  
 This transformation is analogous to the one from Fig.~\ref{fig:blackhole1} to Fig.~\ref{fig:blackhole2} in the black hole case.
The lower sheet in Fig.~\ref{fig:wormhole1} can be similarly deformed into the left-hand side of Fig.~\ref{fig:wormhole3}.
As will be discussed in Sec.\ref{soliton}, the configuration shown in Fig.\ref{fig:wormhole3} naturally emerges when we express the metric in the axisymmetric, conformally flat coordinates~\eqref{eq:ds2static}.

Note that each sheet in Fig.~\ref{fig:wormhole3} is disconnected along the line segment between the points $\bullet$ and $\circ$; we refer to this segment as the \textit{cut}.
As can be seen from Fig.~\ref{fig:wormhole1}, the upper side of the cut on the right sheet in Fig.~\ref{fig:wormhole3} was originally connected to the corresponding part on the left sheet.
The same applies to the lower sides of the cuts.
The presence of these cuts is the primary distinction between black hole and wormhole geometries.
The solitonic configurations incorporating these cuts are constructed in Sec.~\ref{soliton}.

Due to the cuts on the $(\rho, z)$-plane shown in Fig.~\ref{fig:wormhole3}, one might suspect that points on the cut are singular in the $(\rho, z)$ representation.
However, regularity cannot be properly assessed on a disconnected coordinate plane.
Thus, we must return to a connected coordinate system in the neighborhood of such points.
Details are discussed in Sec.~\ref{regularity}.

\subsection{Solitons for wormhole} \label{soliton}

An analytical wormhole solution with a phantom scalar field is presented in Ref.~\cite{Picon2002}:
\begin{equation}
    \dd{s}^2 = -e^{-\frac{2m}{\tilde{r}}}\dd{t}^2 + e^{\frac{2m}{\tilde{r}}}\left(\frac{\alpha/\tilde{r}}{\sin(\alpha/\tilde{r})}\right)^4 \dd{\tilde{r}}^2 + e^{\frac{2m}{\tilde{r}}}\left(\frac{\alpha/\tilde{r}}{\sin(\alpha/\tilde{r})}\right)^2 \tilde{r}^2 \dd{\Omega}^2,
    \label{referencesol}
\end{equation}
with the configuration of the phantom scalar field given by
\begin{equation}
    \Phi = i\frac{Q}{\tilde{r}},
\end{equation}
where $\dd{\Omega}^2$ is the metric of the unit two-sphere, $\dd{\Omega}^2 \coloneqq \dd{\theta}^2 + \sin^2\theta \dd{\phi}^2$, and $i$ denotes the imaginary unit\footnote{
A free phantom scalar field can be represented by taking the scalar field to be purely imaginary in the canonical free scalar field theory.
}.
The radial coordinate $\tilde{r}$, defined in the range $0 < \alpha/\tilde{r} < \pi$, covers the entire spacetime.
This system involves three parameters: the mass $m$, scalar charge $Q$, and a characteristic length $\alpha$.
The Einstein equation
requires 
\begin{equation}
\alpha^2 = Q^2 - m^2.
\label{AlmQ}
\end{equation}
Let us express the metric in the axisymmetric coordinates \eqref{eq:ds2static}. The first step is separating the metric components into the symmetry directions $x^{a} = (t, \phi)$ and the others, thus, 
\begin{align}
   \dd{s}^2 = g_{ab} \dd{x}^{a} \dd{x}^{b} + \dd{l}^2,
\end{align}
with
\begin{align}
 g_{ab} \dd{x}^{a} \dd{x}^{b} &=  -e^{-\frac{2m}{\tilde{r}}}\dd{t}^2 + e^{\frac{2m}{\tilde{r}}}\left(\frac{\alpha \sin \theta}{\sin(\alpha/\tilde{r})}\right)^2  \dd{\phi}^2, \\
\dd{l}^2 &= e^{\frac{2m}{\tilde{r}}}\left(\frac{\alpha/\tilde{r}}{\sin(\alpha/\tilde{r})}\right)^2\left[\left(\frac{\alpha/\tilde{r}}{\sin(\alpha/\tilde{r})}\right)^2\dd{\tilde{r}}^2 + \tilde{r}^2\dd{\theta}^2\right].
\end{align}

Then, following the strategy presented in the previous section,
 $\dd{l}^2$ can be expressed as
\begin{align}
\dd{l}^2 =  F(\rho, z)\qty[\dd{\rho}^2 + \dd{z}^2],    \label{eq:metric3}
\end{align}
in the conformally flat coordinates $(\rho, z)$, which can be introduced as
\begin{equation}
    \left\{ \
    \begin{alignedat}{1}
        \rho &= \sqrt{-\det g_{ab}} = \frac{\alpha\sin\theta}{\sin\qty(\alpha/\tilde{r})} ,\\
        z &= \frac{\alpha\cos\theta}{\tan\qty(\alpha/\tilde{r})} .
    \end{alignedat}
    \right. \label{eq:rhoz1}
\end{equation}

For any point with the coordinate value $(\alpha/\tilde{r}, \theta) \neq (\pi/2, \pi/2)$, there exists the other point with the same $(\rho,z)$ value. Explicitly, a pair of the points with the same $(\rho,z)$ value can be expressed as 
\begin{align}
\left(\frac{\alpha}{\tilde{r}}, \, \theta\right) =  \left(\frac{\alpha}{\tilde{r}_{0}}, \, \theta_{0}\right) \quad \text{and} \quad  \left(\frac{\alpha}{\tilde{r}}, \, \theta\right) = \left(\pi - \frac{\alpha}{\tilde{r}_{0}}, \, \pi - \theta_{0}\right).
\end{align}
Thus, the coordinates $(\rho, z)$ covers only a half of the wormhole geometry and hence we need to introduce two $(\rho,z)$ sheets for describing the entire wormhole geometry.
The upper region of Fig.~\ref{fig:wormhole1} corresponds to the branch with \(0 < \alpha/\tilde{r} < \pi/2\), while the lower region corresponds to that with \(\pi/2 < \alpha/\tilde{r} < \pi\).
The boundary $\alpha/\tilde{r} = \pi/2$, corresponding to $0 \leq \rho \leq \alpha$ with $z = 0$, marks the throat of the wormhole\footnote{The definition of the throat in this paper differs from the standard one, which is defined as a minimal surface. As will be shown in Sec.~\ref{sec:asy}, this wormhole spacetime is \textit{not} symmetric across the throat.}.
Note that the point with $(\alpha/\tilde{r}, \theta) = (\pi/2, \pi/2)$, corresponding to the edge point of the throat $(\rho ,z) = (\alpha, 0)$, does not have the pair. Thus, the points $(\rho ,z) = (\alpha, 0)$ in each sheet must be identified.

Using the \((\rho, z)\) coordinates, \(\tilde{r}\) can be expressed as
\begin{equation}
    \cos\left(\frac{2\alpha}{\tilde{r}}\right) = -\frac{z^2+\alpha^2}{\rho^2} + \frac{\sqrt{\left(\rho^2 + z^2 - \alpha^2\right)^2 + 4\alpha^2 z^2}}{\rho^2} \eqqcolon w. \label{def:w}
\end{equation}
Here, for later convenience, we introduce a variable $w$.
The inverse relation is given by:
\begin{equation}
    \frac{\alpha}{\tilde{r}} 
    = \left\{ \
    \begin{alignedat}{3}
    &\frac{1}{2} \Arccos w \qquad & &\mathrm{for} & \quad 0 &< \frac{\alpha}{\tilde{r}} < \frac{\pi}{2}\\
    &\pi - \frac{1}{2} \Arccos w \qquad & &\mathrm{for} & \quad \frac{\pi}{2} &< \frac{\alpha}{\tilde{r}} < \pi\ .
    \end{alignedat} 
    \right. \label{eq:soliton4}
\end{equation}
Here, $\Arccos$ represents the principal value. Note that
the two branches now appear as the multivaluedness of the inverse trigonometric function.
One can confirm that the function \(\alpha/\tilde{r}\) satisfies the three-dimensional Laplace equation with axial symmetry, and thus it can be interpreted as one of the solitons.  
Accordingly, we introduce solitons \(\mu_{\pm}\) defined by
\begin{equation}
    \left\{ \
    \begin{alignedat}{1}
        \ln \mu_{+} &\coloneqq \Arccos w, \\
        \ln \mu_{-} &\coloneqq 2\pi - \Arccos w,
    \end{alignedat}
    \right. \label{eq:soliton3}
\end{equation}
which satisfy Eq.~\eqref{eq:soliton1}.  
Using Eq.~\eqref{def:w}, one can then express the solitonic forms of the metric functions and the scalar field \(\Phi\) as
\begin{alignat}{2}
    \ln |g_{tt}^{\pm}| &=\ & -\frac{m}{\alpha} &\ln \mu_{\pm}, \label{eq:soliton5gtt}\\
    \ln g_{\phi\phi}^{\pm} &=\ & \frac{m}{\alpha} &\ln \mu_{\pm} + 2\ln \rho, \label{eq:soliton5} \\
    \Phi_{\pm} &=\ & i\frac{Q}{2\alpha} &\ln \mu_{\pm}. \label{eq:soliton5phi}
\end{alignat}

The conformal factor \( F(\rho, z) \) in Eq.~\eqref{eq:metric3} is given by  
\begin{equation}
    F_{\pm}(\rho, z) = (\mu_{\pm})^{\frac{m}{\alpha}} \left[1 - \frac{\rho^2}{4\alpha^2}(1 - w)^2\right]^{-1}. \label{eq:factor1}
\end{equation}
Therefore, we obtain the line element in the form of Eq.~\eqref{eq:ds2static} as  
\begin{equation}
    \dd{s}_{\pm}^2 = -(\mu_{\pm})^{-\frac{m}{\alpha}} \dd{t}^2 + (\mu_{\pm})^{\frac{m}{\alpha}} \rho^2 \dd{\phi}^2 + (\mu_{\pm})^{\frac{m}{\alpha}} \left[1 - \frac{\rho^2}{4\alpha^2}(1 - w)^2\right]^{-1} \qty[\dd{\rho}^2 + \dd{z}^2]. \label{eq:fullmetric}
\end{equation}

To summarize, the spherically symmetric wormhole solutions are now expressed in terms of the solitonic formulation involving two sheets: the metric is given by Eq.~\eqref{eq:fullmetric} and the scalar field is given by Eq.~\eqref{eq:soliton5phi}, where the functions $\mu_{\pm}(\rho, z)$ and $w(\rho,z)$ are defined in Eqs.~\eqref{eq:soliton3} and \eqref{def:w}, respectively.
Hereafter, we omit the plus-minus subscript unless otherwise specified.

\subsection{Regularities} \label{regularity}
The solution~\eqref{referencesol} is manifestly regular even on the throat, since it is represented in the global coordinates $(\tilde{r},\theta)$. 
However, this regularity is not immediately clear in the solitonic form~\eqref{eq:fullmetric}, because the throat appears as a cut in the \((\rho, z)\) sheets.
In this section, we explain how to verify the regularity of the solution within the solitonic constructions.

In addition to confirming the regularity along the axis, as is done in standard solitonic constructions, we must also address the issue of regularity at the junction between two sheets.  
These regions lie on the boundaries of each sheet, meaning that the analysis should be carried out in a coordinate system that connects the two sheets.

The metric components \(g_{tt},\ g_{\phi\phi}\), and the scalar field \(\Phi\) are expressed in terms of the soliton \(\mu\),  
which we define over the entire region such that it equals \(\mu_+\) on one sheet and \(\mu_-\) on the other.  
Thus, verifying the regularity of these fields reduces to analyzing the soliton \(\mu\).  
Moreover, the geometric regularity also depends critically on the conformal factor \(F(\rho, z)\),  
because the flat \((\rho, z)\) sheet shown in Fig.~\ref{fig:wormhole3} is related to the physical geometry through this factor.
In particular, at the edge of the cut (the point marked \(\circ\) in Fig.~\ref{fig:wormhole3}),  
the conformal transformation becomes singular, requiring a detailed analysis of regularity at that point.  
We denote this edge point, \((\rho, z) = (\alpha, 0)\), as \p.

To ensure regularity, the following conditions must be verified:
\begin{enumerate}[label=\textbf{\arabic*.}]
    \item \textbf{Regularity across the cut:} The soliton \(\mu\) and the conformal factor \(F(\rho, z)\) are smoothly connected between the two sheets across the cut \(z = 0,\ 0 < \rho < \alpha\).
    \item \textbf{Regularity of soliton $\mu$ at the edge point $p_{\circ}$:} 
    The soliton \(\mu\) satisfies the covariantly written Laplace equation
    \begin{equation}
        \D^i\qty(\sqrt{-g} \, \D_i \ln \mu) = 0, \label{eq:Lap}
    \end{equation}
    everywhere in the spacetime.
    At the edge point \p, this condition needs to be verified in a regular coordinate system.
    \item \textbf{Regularity of $\dd{l}^2$ at the edge point $p_{\circ}$:} The edge point \p does not introduce a conical singularity. That is, the spatial line element,
    \begin{equation}
        \dd{l}^2 = F(\rho, z)\left[\dd{\rho}^2 + \dd{z}^2\right],
    \end{equation}
    is regular at \p.
    \item \textbf{Regularity on the axis:} The axis \(\rho = 0\) does not introduce a conical singularity. The line element along the axis,
    \begin{equation}
        \dd{s}^2\big|_{\text{axis}} = F(\rho, z)\dd{\rho}^2 + g_{\phi\phi}(\rho,z)\dd{\phi}^2 + g_{tt}(\rho,z) \dd{t}^2 + F(\rho,z) \dd{z}^2,
    \end{equation}
    remains regular.
    Note that only the first two terms are potentially involving a conical singularity.
\end{enumerate}

In addition to these, we impose asymptotic flatness on each sheet:
\begin{enumerate}
    \item[{\bf 5.}] \textbf{Asymptotic flatness:} Each asymptotic region is asymptotically flat.
\end{enumerate}

The first three conditions are specific to wormhole construction, while the other two are standard in solitonic approaches (see Refs.~\cite{Belinski1978,Belinski1979,Vigano:2022hrg}). 
Let us check each condition in detail.

\subsubsection{Regularity across the cut} 

The first condition ensures that both $\mu$ and $F$ satisfy the Laplace equation~\eqref{eq:Lap} and the Einstein equations across the cut.  
The soliton $\mu$ is constructed to solve Eq.~\eqref{eq:Lap} in the $(\rho,z)$ coordinates, except at the boundaries of each sheet.  
Since within the interior of each sheet there exists a regular transformation from a smooth general coordinate system that is valid over the whole domain to the $(\rho,z)$ coordinates,  
the function $\mu$ also solves Eq.~\eqref{eq:Lap} in the general coordinates, due to the conformal invariance of Eq.~\eqref{eq:Lap}.  
Because Eq.~\eqref{eq:Lap} is a second-order quasi-linear differential equation, $C^1$ continuity of $\mu$ across the cut ensures that Eq.~\eqref{eq:Lap} is satisfied there.  
Similarly, since the conformal factor $F$ appears with derivatives up to first order in the Einstein equations, continuity of $F$ guarantees that the equations are satisfied.  
Therefore, the smoothness of both $\mu$ and $F$ is sufficient to ensure they are valid solutions on the  cut.

\begin{figure}[t]
    \centering
    \includegraphics{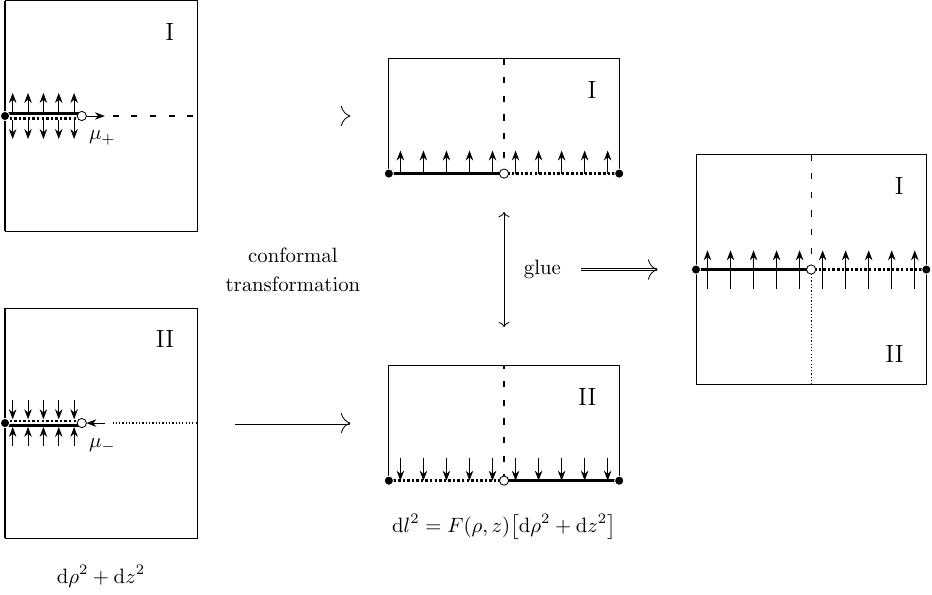}
    \caption{Gluing of two sheets. The arrows on the sheets indicate the gradient field of the solitons, which is discontinuous within each sheet (see left panels) but becomes continuous once the sheets are glued together (see right panel). The edge of the cuts, $p_{\circ}$, appears singular with respect to the flat metric $\dd{\rho}^2 + \dd{z}^2$, but is regular in terms of the physical metric $\dd{l}^2$ after gluing.
    }
    \label{fig:grad1}
\end{figure}

Let us verify that the first condition is indeed satisfied in the solution~\eqref{eq:fullmetric}.
First, $\rho$ is trivially smooth across the cut, because the both cuts lie along the segment where $z = 0$ and $0 \le \rho < \alpha$.
Second, let us consider the behavior of $w$, defined in Eq.~\eqref{def:w}.  
The cut in each sheet corresponds to $0 \le \rho < \alpha$ and $z = 0$.  
Along this cut, the expression under the square root in Eq.~\eqref{def:w} is always strictly positive,  
and thus $w$ is smooth.
Third, from Eq.~\eqref{eq:soliton3} and the fact that the cut occurs where $\ln \mu = \pi$, $\mu$ is clearly a smooth function of $w$ from the viewpoint of the covering space of $\Arccos w$, namely $\arccos w$.
Although the regularity of $\ln \mu$ across the cut has already been verified, 
examining the expansions of these functions may provide a more direct understanding of the regularity.
The function $\Arccos{w}$ is constant along the cut $z=0, \ 0\leq\rho<\alpha$ because it corresponds to $\alpha/\tilde{r} = \pi/2$. Since the function $w$ can be expanded around $z\rightarrow0$ as
\begin{equation}
    w = -1 + \frac{1}{2\rho^2}\frac{\alpha^2+3\rho^2}{\alpha^2-\rho^2}z^2 + \order{z^4},
\end{equation}
we obtain the asymptotic behavior of the solitons $\mu$ as
\begin{equation}
    \ln\mu_{\pm} = \pi \mp \sqrt{\frac{\alpha^2+3\rho^2}{\alpha^2-\rho^2}}\frac{\abs{z}}{\rho} + \order{z^2}. \label{eq:mu1}
\end{equation}
It implies that, within a single sheet, the solitons $\mu=\mu_{\pm}$ exhibit a cusp along the cut as shown in the left panels of Fig.~\ref{fig:grad1}. 
Meanwhile, if one sheet is glued to the other, $\mu$ has at least $C^1$ continuity as shown in the right panel of Fig.~\ref{fig:grad1}. 

\subsubsection{Regularity of soliton $\mu$ at the edge point \p } \label{regcut}

Regularity at the edge point \p can be verified by expanding the relevant functions around \p and confirming that the soliton $\mu$ satisfies the Laplace equation~\eqref{eq:Lap} directly. 
The equation for solitons, namely Eq.~\eqref{eq:soliton1}, is expressed in the covariant form~\eqref{eq:Lap} in general coordinates~\eqref{eq:dl}. 
At the edge \p of the cut, since the conformally flat coordinates $(\rho, z)$ become singular, a regular coordinate system needs to be introduced.

Before introducing a regular coordinate system, we note the conformal invariance of Eq.~\eqref{eq:Lap}.  
Consider a conformal transformation
\begin{equation}
\gamma_{ij}(y^k) =: \Omega^2(y^k)\, \tilde{\gamma}_{ij}(y^k),
\end{equation}
where $\Omega(y^k)$ is nonzero and finite everywhere.  
Even after this transformation, the form of Eq.~\eqref{eq:Lap} remains invariant, i.e., it is equivalent to
\begin{equation}
\tilde{\D}^i\qty(\sqrt{-g} \, \tilde{\D}_i \ln \mu) = 0, \label{eq:confLap}
\end{equation}
where $\tilde{\D}_{i}$ denotes the covariant derivative with respect to $\tilde{\gamma}_{ij}$.
Therefore, the conformal factor can be ignored as long as it remains regular.

\begin{figure}[t]
    \centering
    \includegraphics{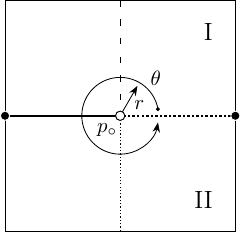}
    \caption{A polar coordinate system $(r,\theta)$ around the point \p. When the two sheets are combined, the angle coordinate $\theta$ runs from $0$ to $2\pi$. The coordinate system~\eqref{eq:2drtheta} around \p is regular  if the conformal factor $4 F r^2$ remains finite and non-zero in the limit $r \to 0$.}
    \label{fig:entire3}
\end{figure}

We introduce polar coordinates \((r, \theta)\) around the point \p via the coordinate transformation
\begin{equation}
    \left\{
    \begin{alignedat}{1}
        \rho - \alpha &= -r^2\cos 2\theta, \\
        z &= -r^2\sin 2\theta,
    \end{alignedat}
    \right.
    \label{rtheta2}
\end{equation}
where the coordinates \((r, \theta)\) are defined by \(r \geq 0\) and
\begin{equation}
    \left\{
    \begin{alignedat}{2}
        0 & \leq \theta < \pi &\qquad& \text{on sheet \textrm{I}}, \\
        \pi & \leq \theta < 2\pi &\qquad& \text{on sheet \textrm{II}},
    \end{alignedat}
    \right.
\end{equation}
spanning both sheets around the edge point \p. 
The line element on the connected sheet becomes
\begin{equation}
    \dd{l}^2 = F(\rho, z)\qty[\dd{\rho}^2 + \dd{z}^2] = 4Fr^2\qty[\dd{r}^2 + r^2\dd{\theta}^2].
    \label{eq:2drtheta}
\end{equation}
We will show in Sec.~\ref{sec:regp} that this coordinate system is regular around \p, i.e., the conformal factor $4Fr^2$ remains finite around \p.
The coordinate configuration is illustrated in Fig.~\ref{fig:entire3}. 
Then, by conformal invariance, the Laplace equation~\eqref{eq:Lap} can be expressed in the polar coordinates of the two-dimensional flat space; we denote its covariant derivative by $\bar \D_i$.

The asymptotic behavior as \(r \to 0\) is given by
\begin{align}
    w &= -1 + 4\left(\frac{r^2}{\alpha}\right)\sin^2\theta + \order{\frac{r^4}{\alpha^2}}, \\
    \ln \mu_{\pm} &= \pi \pm 2\sqrt{2}\left(\frac{r\left|\sin \theta\right|}{\sqrt{\alpha}}\right) + \order{\frac{r^3}{\alpha ^{3/2}}}. \label{eq:scalar3}
\end{align}
The solitons \(\mu_\pm\) can be expressed together as
\begin{align}
    \ln \mu = \pi + 2\sqrt{2}\left(\frac{r \sin \theta}{\sqrt{\alpha}}\right) + \order{\frac{r^3}{\alpha ^{3/2}}}.
    \label{eq:expm}
\end{align}
Recalling that \(\sqrt{-g} = \rho\) and using Eq.~\eqref{rtheta2},  
the Laplace equation~\eqref{eq:Lap} in the polar coordinate of the flat space can be rewritten as
\begin{equation}
\left(\alpha - r^2\cos 2\theta\right)  {\bar \D}^i{\bar \D}_i \ln \mu 
 - \left[{\bar \D}^i \qty(r^2\cos 2\theta)\right]{\bar \D}_i \ln \mu = 0.
\label{eq:Lapex}
\end{equation}
From the expansion of \(\ln \mu\) in Eq.~\eqref{eq:expm}, we find
\begin{align}
\lim_{r \to 0} \bar\D^i \bar\D_i \ln \mu &= 0, \\
\lim_{r \to 0} \left|\bar\D_i \ln \mu\right| &< \infty.
\end{align}
The first equation shows that the Laplacian term, i.e., the first term on the left-hand side of Eq.~\eqref{eq:Lapex}, vanishes as \(r \to 0\),  
while the second ensures that the derivative term remains finite,  
which implies that the second term on the left-hand side of Eq.~\eqref{eq:Lapex} also vanishes.  
Therefore, \(\ln \mu\) satisfies the Laplace equation~\eqref{eq:Lap} at the edge point \p.

We also verify that $\ln \rho$ satisfies the Laplace equation~\eqref{eq:Lap}, as expected from its definition~\eqref{eq:gauge1} and Eq.~\eqref{eq:gharm}. 
At \p, $\ln \rho$ can be expanded as
\begin{equation}
\ln \rho = \ln \left( \alpha - r^2 \cos 2\theta \right)
= \ln \alpha - \frac{r^2}{\alpha} \cos 2\theta + \order{\frac{r^4}{\alpha^2}}.
\end{equation}
Substituting $\ln \mu = \ln \rho$ into Eq.~\eqref{eq:Lapex} shows that the Laplace equation~\eqref{eq:Lap} indeed holds for $\ln \rho$ in the limit $r \to 0$.

\subsubsection{Regularity of $\mathrm{d}{l}^2$ at the edge point \p} 
\label{sec:regp}

From the viewpoint of the $(\rho,z)$ sheets shown in the left panels of Fig.~\ref{fig:grad1}, the two sheets are glued together along the cuts, and the edge point \p of the cut corresponds to
a conical singularity with respect to the flat metric $\dd{\rho}^2 + \dd{z}^2$.  
However, the original geometry is smooth, as illustrated in Fig.~\ref{fig:wormhole1}. This apparent discrepancy arises because the conformal factor $F$ is singular at \p and the physical metric $\dd{l}^2 = F \qty[\dd{\rho}^2 + \dd{z}^2]$ is regular (see the right panel of Fig.~\ref{fig:grad1}). 
Let us see this structure in detail.

Under the coordinate transformation~\eqref{rtheta2},  
the line element of the $(\rho,z)$ plane near the edge point \p of the cut takes the form~\eqref{eq:2drtheta}.
Recall that the coordinate $\theta$ ranges from $0$ to $2\pi$ over the entire space, covering both sheets.
From the explicit form of $F$ in Eq.~\eqref{eq:factor1},  
the asymptotic behavior of $F(\rho, z)$ in the limit $r \rightarrow 0$ is
\begin{equation}
    F \approx e^{\frac{\pi m}{\alpha}}\frac{\alpha}{2r^2} \propto \frac{1}{r^2}, \label{eq:asympF1}
\end{equation}
where $\approx$ means that the higher order in $r$ is ignored.
Thus, the line element near the edge point \p becomes
\begin{equation}
    \dd{l}^2 \approx 2\alpha\, e^{\frac{\pi m}{\alpha}} \qty[\dd{r}^2 + r^2 \dd{\theta}^2].
\end{equation}
This indicates that no conical singularity arises at the point \p.

\subsubsection{Regularity on the axis} 
\label{regax}
The geometry on the axis can be understood by fixing $t$ and $\theta$ in the line element~\eqref{eq:fullmetric},
\begin{equation}
    \dd{s}^2\Big|_{\text{axis}} = \mu^{\frac{m}{\alpha}}\qty{\qty[1 - \frac{\rho^2}{4\alpha^2}\qty(1 - w)^2]^{-1}\dd{\rho}^2 + \rho^2\dd{\phi}^2} + (\text{regular terms}).
\end{equation}
Taking the limit $\rho \rightarrow 0$ gives the following results:
\begin{align}
    w &= \frac{z^2 - \alpha^2}{z^2 + \alpha^2} + \order{\rho^2}, \\[1mm]
    \mu_{+} &= \underbrace{\exp\qty[\Arccos\qty(\frac{z^2 - \alpha^2}{z^2 + \alpha^2})]}_{\eqqcolon \mu_{0}^{+}} + \order{\rho^2}, \\
    \mu_{-} &= \underbrace{e^{2\pi}\exp\qty[-\Arccos\qty(\frac{z^2 - \alpha^2}{z^2 + \alpha^2})]}_{\eqqcolon \mu_{0}^{-}} + \order{\rho^2}.
\end{align}
Therefore, the asymptotic form of the line element near the axis is given by
\begin{equation}
    \dd{s}_{\pm}^2\Big|_{\text{axis}} \approx \qty(\mu_{0}^{\pm})^{\frac{m}{\alpha}}\qty[\dd{\rho}^2 + \rho^2\dd{\phi}^2] + \text{(regular terms)}.
\end{equation}
This shows that no conical singularity appears on the axis, since the azimuthal angle $\phi$ ranges from $0$ to $2\pi$, and the prefactor $\mu_0^\pm$ remains constant with respect to $\rho$.

\subsubsection{Asymptotic flatness} \label{sec:asy}

The asymptotic region corresponds to the limit in which either $\rho$ or $z$ tends to infinity. Given that the intrinsic scale of the wormhole is the coordinate value $\alpha$ of the throat, this limit can be characterized by the parameter
\begin{equation}
 \epsilon:= \frac{\alpha}{\sqrt{\rho^2+z^2}} \ \longrightarrow \ 0.
 \label{limrz}
\end{equation}
In the sheet associated with $\mu_+$, the functions $\mu_+$, $w$, $g_{tt}$, $g_{\phi\phi}$, and $F$ admit expansions in powers of $\epsilon$ as follows:
\begin{align}
    \mu_{+} &= 1 + \frac{2\alpha}{\sqrt{\rho^2+z^2}} + \order{\epsilon^2}, \\
    w &= 1 - \frac{2\alpha^2}{\rho^2+z^2} + \order{\epsilon^4}, \\
    g_{tt}^{+} &= -\qty(1 - \frac{2m}{\sqrt{\rho^2+z^2}}) + \order{\epsilon^2}, \label{AE+} \\
    g_{\phi\phi}^{+} &= \rho^2\qty(1 + \frac{2m}{\sqrt{\rho^2+z^2}}) + \order{\epsilon^2}, \\
    F &= 1 + \frac{2m}{\sqrt{\rho^2+z^2}} + \order{\epsilon^2}.
\end{align}
Upon introducing the coordinate transformation
\begin{equation}
\rho =\rp\sin \theta, \qquad z=\rp\cos \theta, \label{AStrans+}
\end{equation}
the line element~\eqref{eq:fullmetric} asymptotically becomes
\begin{equation}
\dd{s}^2 \approx - \left( 1-\frac{2m}{\rp} \right) \dd{t}^2 + 
 \left( 1+\frac{2m}{\rp} \right)\qty\Big[\dd{\rp}^2 +\rp^2 \dd{\theta}^2 + \rp^2 \sin^2\theta \dd{\phi}^2].
\end{equation}
This indicates that the asymptotic region of the $\mu_+$ sheet is asymptotically flat with ADM mass $m$.

A similar analysis can be carried out for the sheet associated with $\mu_-$. The functions $\mu_-$, $w$, $g_{tt}$, $g_{\phi\phi}$, and $F$ admit expansions in $\epsilon$ of the form,
\begin{align}
    \mu_{-} &= e^{2\pi}\qty(1 - \frac{2\alpha}{\sqrt{\rho^2+z^2}}) + \order{\epsilon^2}, \\
    w &= 1 + \frac{2\alpha^2}{\rho^2+z^2} + \order{\epsilon^4}, \\
    g_{tt}^{-} &= -e^{-\frac{2\pi m}{\alpha}}\qty(1 + \frac{2m}{\sqrt{\rho^2+z^2}}) + \order{\epsilon^2}, \label{AE-} \\
    g_{\phi\phi}^{-} &= e^{\frac{2\pi m}{\alpha}}\rho^2\qty(1 - \frac{2m}{\sqrt{\rho^2+z^2}}) + \order{\epsilon^2}, \\
    F &= e^{\frac{2\pi m}{\alpha}} \qty(1 - \frac{2m}{\sqrt{\rho^2+z^2}}) + \order{\epsilon^2}.
\end{align}
By applying the coordinate transformation,
\begin{equation}
t= e^{-\frac{\pi m}{\alpha}}t_{(-)}, \qquad \rho = e^{\frac{\pi m}{\alpha}}\rn\sin \theta, \qquad z=e^{\frac{\pi m}{\alpha}}\rn\cos \theta, \label{AStrans-}
\end{equation}
the metric~\eqref{eq:fullmetric} becomes
\begin{equation}
\dd{s}^2 \approx - \left( 1+\frac{2me^{-\frac{\pi m}{\alpha}}}{\rn} \right) \dd{t_{(-)}}^2 + \left( 1-\frac{2me^{-\frac{\pi m}{\alpha}}}{\rn} \right)\qty\Big[\dd{\rn}^2 +\rn^2 \dd{\theta}^2 + \rn^2 \sin^2\theta \dd{\phi}^2].
\end{equation}
Hence, the sheet associated with $\mu_-$ is also asymptotically flat, with ADM mass $-me^{-\pi m/\alpha}$.

\section{Multi-sheet wormhole solutions} \label{multisheet}

In this section, we generalize the construction of two-sheet wormhole solutions presented in the previous section to obtain multi-sheet wormhole solutions where a wormhole connects multiple asymptotic regions.

\begin{figure}[t]
    \centering
    \includegraphics{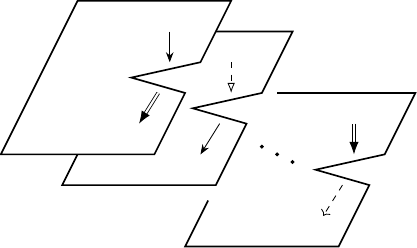}
    \caption{An illustration of the construction of a {\it multi-sheet} wormhole. Each sheet corresponds to a single $(\rho,z)$ sheet with a cut. The geometry is constructed by gluing these sheets across the cuts marked with the same arrow.}
    \label{fig:sheet2}
\end{figure}

The key idea for constructing a wormhole solution that connects multiple asymptotic regions is to prepare multiple $(\rho,z)$ sheets and glue them together in the manner illustrated in Fig.~\ref{fig:sheet2}.
A notable feature is that the cut structure required to realize Fig.~\ref{fig:sheet2} remains the same as that of the two-sheet wormhole, and we now see that it can be realized using the solitons $\mu_{\pm}$ obtained in the previous section.

In Sec.~\ref{MSsoliton}, we introduce the solitons $\mu_{\pm}$ on each sheet and present a four parameter family of metric and scalar field configurations, characterized by $m$, $Q$, $\alpha$, and $A$, using the soliton $\mu_{\pm}$. 
Finally, in Sec.~\ref{MSreg}, we examine the regularity of the solutions, which yields a constraint among these parameters.

\subsection{Solutions on each sheet} \label{MSsoliton}
Based on the key idea explained above, we assume that the metric and scalar field configuration are expressed in the same form as given in Eqs. \eqref{eq:soliton5gtt} to \eqref{eq:soliton5phi}, thus,
\begin{alignat}{2}
    \ln |g_{tt}^{\pm}| &=\ & -\frac{m}{\alpha} &\ln \mu_{\pm}(\rho,z), \label{eq:soliton6gtt}\\
    \ln g_{\phi\phi}^{\pm} &=\ & \frac{m}{\alpha} &\ln \mu_{\pm}(\rho,z) + 2\ln \rho, \label{eq:soliton6}\\
    \Phi_{\pm} &=\ & i\frac{Q}{2\alpha} &\ln \mu_{\pm}(\rho,z). \label{eq:soliton6phi}
\end{alignat}
Here the parameters $m, Q$ and $\alpha$ are introduced as free parameters and the condition \eqref{AlmQ} is not necessarily imposed. 
By construction, Eqs.~\eqref{eq:soliton6gtt} to \eqref{eq:soliton6phi} are solutions of the Laplace equations. The remaining components of the Einstein equation, Eqs.~\eqref{eq:eqforFstatic}, determine the conformal factor $F$. In the absence of the condition \eqref{AlmQ}, the conformal factor is modified from that in Eq.~\eqref{eq:factor1}, and given as 
\begin{equation}
    F_{\pm}(\rho, z) = A \left(\mu_{\pm}\right)^{\frac{m}{\alpha}} \left[1 - \frac{\rho^2}{4\alpha^2}(1 - w)^2\right]^{-B},
    \label{MSF}
\end{equation}
where $A$ is an integration constant, and $B$ is a constant given by
\begin{equation}
    B \coloneqq \frac{Q^2 - m^2}{\alpha^2}.
    \label{defB}
\end{equation}
It is worth noting that when the condition \eqref{AlmQ} is satisfied, the exponent $B$ reduces to unity, and Eq.~\eqref{eq:factor1} is recovered by choosing $A = 1$.

At this stage, we obtain two choices of the solution on each sheet,
\begin{equation}
    \dd{s}_{\pm}^2 = -(\mu_{\pm})^{-\frac{m}{\alpha}} \dd{t}^2 + (\mu_{\pm})^{\frac{m}{\alpha}} \rho^2 \dd{\phi}^2 + A \, (\mu_{\pm})^{\frac{m}{\alpha}} \left[1 - \frac{\rho^2}{4\alpha^2}(1 - w(\rho,z))^2\right]^{-B} \qty[\dd{\rho}^2 + \dd{z}^2], \label{eq:fullmetricB}
\end{equation}
along with the corresponding scalar field given in Eq.~\eqref{eq:soliton6phi}. 
The period of the coordinate $\phi$ can be arbitrary at this stage, but we set it to $2\pi$.\footnote{
If the period of $\phi$ is not $2\pi$, rescaling $\phi$ as $\tilde \phi := b \phi$ with a constant $b$ can make the period of the new coordinate $\tilde \phi$ equal to $2\pi$.
Then, by appropriately rescaling $\rho$, $z$, $m$, $\alpha$, $Q$ and $A$, the line element and the scalar field take the same form as the original ones.
Hence, without loss of generality, the period of $\phi$ can be set to $2\pi$.
}

\subsection{Gluing Sheets and Regularities}\label{MSreg}

The remaining task is to assign one of the solutions \eqref{eq:fullmetricB} to each sheet and adjust the parameters in such a way that the entire spacetime, obtained by gluing all sheets together as in Fig.~\ref{fig:sheet2}, becomes regular.

\begin{figure}[t]
    \centering
    \includegraphics{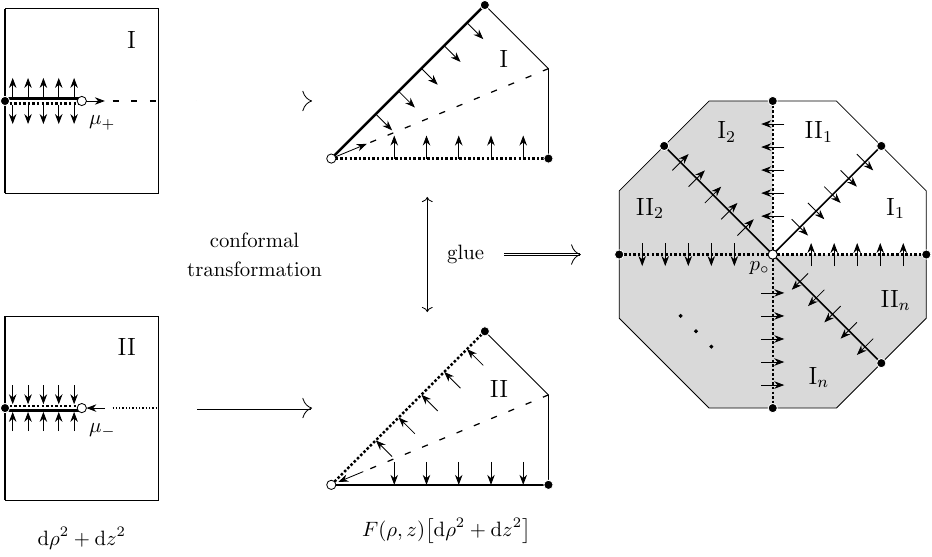}
    \caption{A {\it multi-sheet} generalization of the procedure shown in Fig.~\ref{fig:grad1}. The gradient field of the solitons becomes continuous again once the sheets, $\mathrm{I}_{k}$ and $\mathrm{II}_{k}$, are glued alternately (see the right panel).}
    \label{fig:entire2}
\end{figure}

In the following section, we verify all the regularity conditions, as well as the asymptotically flat conditions, summarized in Sec.~\ref{regularity}.

\subsubsection{Regularity across the cut}
As is evident from the analysis of two-sheet case, the soliton $\mu_+$ must be smoothly connected to $\mu_-$ across all cuts. 
This can be achieved by preparing an even number of sheets and assigning $\mu_+$ and $\mu_-$ alternately. Accordingly, we adopt this configuration.
As Fig.~\ref{fig:grad1} for the two-sheet wormhole, the structure near the edge point can be represented as in Fig.~\ref{fig:entire2}.
Half of these sheets, denoted as $\mathrm{I}_k$ ($1 \leq k \leq n$), host the soliton $\mu_+$, 
while the remaining half, labeled as $\mathrm{II}_k$ ($1 \leq k \leq n$), contain the soliton $\mu_-$. 
Due to the behavior of the conformal factor $F$, each sheet resembles a slice of pizza, and the entire configuration formed by gluing them together corresponds to a complete pizza (see Fig.~\ref{fig:entire2}).
Let the number of sheets be denoted by $2n$, and define the soliton $\mu$ such that it coincides with $\mu_+$ on $\mathrm{I}_k$ and with $\mu_-$ on $\mathrm{II}_k$, respectively.

Since this local structure is identical to that of the two-sheet wormhole case, the smoothness of each function is ensured. 
Although certain exponents in the function $F$ may differ from those in the two-sheet configuration, this variation does not affect the regularity on the cuts. Therefore, all functions remain regular across the cuts.

\subsubsection{Regularity of soliton $\mu$ at the edge point \p}

\begin{figure}[t]
    \centering
    \includegraphics{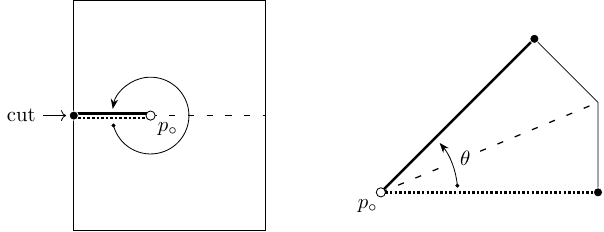}
    \caption{An example of a conformal transformation. As shown in Fig.~\ref{fig:grad1}, the edge \p is singular with respect to the flat metric $\dd{\rho}^2+\dd{z}^2$. The appropriate gluing of the sheets will eliminate this singularity in terms of the physical metric $\dd{l}^2$.}
    \label{fig:conformal}
\end{figure}

To examine the regularity of soliton $\mu$ at the edge \p of the cut, 
we analyze its behavior from the perspective of the right-hand diagram in Fig.~\ref{fig:entire2}. 
Let us introduce the coordinate transformation
\begin{equation}
    \left\{
    \begin{alignedat}{1}
        \rho - \alpha &= -r^{2n} \cos\qty(2n\theta), \\
        z &= -r^{2n} \sin\qty(2n\theta),
    \end{alignedat}
    \right. \label{MStrans}
\end{equation}
with the angular domain specified as
\begin{equation}
\left\{
    \begin{alignedat}{2}
        \frac{2(k-1)}{n}\pi & \leq \theta < \frac{2k-1}{n}\pi &\qquad& \text{on sheet $\textrm{I}_{k}$}, \\
        \frac{2k-1}{n}\pi & \leq \theta < \frac{2k}{n}\pi &\qquad& \text{on sheet $\textrm{II}_{k}$}.
    \end{alignedat}
\right. \label{MSangle}
\end{equation}
This transformation leads to the line element
\begin{equation}
    \dd{l}^2 = F(\rho, z)\qty[\dd{\rho}^2 + \dd{z}^2] 
    = 4n^2 F r^{4n-2} \qty[\dd{r}^2 + r^2 \dd{\theta}^2]. 
    \label{eq:conformal1}
\end{equation}
As will be shown in Sec.~\ref{MSCon}, this expression can be regular at $r=0$. 
The $(r, \theta)$ coordinate system is illustrated in Fig.~\ref{fig:conformal}.

To check the regularity of the solitons $\mu$, we should examine that Eq.\eqref{eq:Lap} is satisfied at \p. 
The asymptotic behavior of soliton $\mu_\pm$ as $r\to0$ becomes
\begin{equation}
    \ln \mu_{\pm} = \pi \pm 2\sqrt{2}\qty(\frac{r^{n}\left|\sin\qty(n\theta)\right|}{\alpha}) +\order{\frac{r^{3n}}{\alpha^{3/2}}}.
\end{equation}
Recalling the range of $\theta$ on each sheet, $\mu$ is written as
\begin{equation}
    \ln \mu = \pi +2\sqrt{2}\qty(\frac{r^{n} \sin \qty(n\theta)}{\sqrt{\alpha}}) + \order{\frac{r^{3n}}{\alpha^{3/2}}}.
\end{equation}
This implies that the soliton equation~\eqref{eq:Lap} is satisfied in the limit $r\to0$.

At \p, $\ln \rho$ can be expanded as
\begin{equation}
    \ln \rho = \ln\alpha - \frac{r^{2n}}{\alpha} \cos\qty(2n\theta) 
    + \order{\frac{r^{4n}}{\alpha^2}}.
\end{equation}
By substituting $\mu = \rho$ into Eq.~\eqref{eq:Lapex}, one finds that this expression also satisfies the Laplace equation~\eqref{eq:Lap}.

\subsubsection{Regularity of $\mathrm{d}{l}^2$ at the edge point \p} 
\label{MSCon}

\begin{figure}[t]
    \centering
    \includegraphics{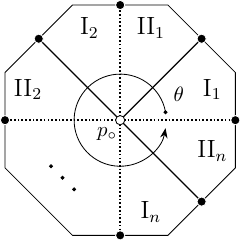}
    \caption{The angle coordinate around \p. Each piece covers an angle of $\pi/n$. Gluing $2n$ of them together yields the complete structure.}
    \label{fig:coord1}
\end{figure}

The physical metric $\dd{l}^2$ is now expressed as Eq.~\eqref{eq:conformal1}.
As indicated in Eq.~\eqref{MSangle}, the angular coordinate $\theta$ spans a domain of width $\pi/n$ on each individual sheet. Consequently, it ranges from $0$ to $2\pi$ over the entire multi-sheeted structure composed of $2n$ sheets (see Fig.~\ref{fig:coord1}).

The metric form~\eqref{eq:metric4} implies that the regularity around the point \p depends on the behavior of the conformal factor $F$ in $r \rightarrow 0$ limit. 
Following the same analysis presented in Section~\ref{regularity}, the conformal factor $F$, given in Eq.~\eqref{MSF}, can be expanded in powers of $r$ as
\begin{equation}
    F \approx A\, e^{\frac{\pi m}{\alpha}} \qty(\frac{\alpha}{2r^{2n}})^{B} \propto \frac{1}{r^{2nB}}.
\end{equation}
Therefore, $\dd{l}^2$ can be expanded as 
\begin{equation}
    \dd{l}^2 \approx (\text{constant}) \times r^{2(2n-1- n B)} \qty[\dd{r}^2 + r^2 \dd{\theta}^2]. \label{eq:metric4}
\end{equation}
Hence, the regularity around \p is ensured if and only if the condition $B = \qty(2n-1)/n$ is satisfied. From Eq.~\eqref{defB}, this implies that the square difference between the charge $Q$ and mass $m$ is determined by the number of the sheets,
\begin{equation}
    Q^2 - m^2 = \frac{2n - 1}{n} \alpha^2.
    \label{MSmQal}
\end{equation}

\subsubsection{Regularity on the axis} 
\label{MSaxis}

A similar analysis to that in Section~\ref{regax} yields the expansion of the metric near the axis as
\begin{equation}
    \dd{s}_{\pm}^2|_{\text{axis}} \approx \qty(\mu_{0}^{\pm})^{\frac{m}{\alpha}} \qty[A\dd{\rho}^2 + \rho^2\dd{\phi}^2] + \text{(regular terms)}.
\end{equation}
Accordingly, regularity on the axis is ensured if and only if the condition $A = 1$ is satisfied. 
As will be discussed in Section~\ref{MSAF}, this regularity condition $A = 1$ also guarantees asymptotic flatness.

\subsubsection{Asymptotic flatness} \label{MSAF}

The asymptotic region can be analyzed following the same procedure as in Section~\ref{sec:asy}, 
namely by taking the limit~\eqref{limrz}. 
The metric functions and scalar field are expanded as in Eqs.~\eqref{AE+} and \eqref{AE-}; 
however, since the form of $F$ differs from that in the two-sheeted wormhole case (see Eq.~\eqref{MSF}), 
the conformal factor $F$ takes the following form:
\begin{align}
    F_{+} (\rho,z) &= A \qty(1 + \frac{2m}{\sqrt{\rho^2 + z^2}}) + \order{\epsilon^2}, \\
    F_{-} (\rho,z) &= A\, e^{\frac{2\pi m}{\alpha}} \qty(1 - \frac{2m}{\sqrt{\rho^2 + z^2}}) + \order{\epsilon^2}.
\end{align}
Although the constant $A$ has been fixed to unity in Sec.~\ref{MSaxis}, in order to demonstrate that this result 
also follows from the condition of asymptotic flatness, we do not impose $A = 1$ at this stage.

By adopting the same coordinate transformations as those used for the two-sheeted wormhole (Eqs.~\eqref{AStrans+} and \eqref{AStrans-}), 
the asymptotic forms of the regions $\mathrm{I}_k$ and $\mathrm{II}_k$ are given by
\begin{align}
    \dd{s}_{\mathrm{I}}^2 &\approx - \left( 1-\frac{2m}{\rp} \right) \dd{t}^2 + \left( 1+\frac{2m}{\rp} \right)\qty\Big[A\dd{\rp}^2 + A\,\rp^2 \dd{\theta}^2 + \rp^2 \sin^2\theta \dd{\phi}^2], \\
    \dd{s}_{\mathrm{II}}^2 &\approx - \left( 1+\frac{2me^{-\frac{\pi m}{\alpha}}}{\rn} \right) \dd{t_{(-)}}^2 + \left(1-\frac{2me^{-\frac{\pi m}{\alpha}}}{\rn} \right)\qty\Big[A\dd{\rn}^2 + A\,\rn^2 \dd{\theta}^2 + \rn^2 \sin^2\theta \dd{\phi}^2],
\end{align}
respectively. These spacetimes are asymptotically flat if and only if the condition $A = 1$ is satisfied.  
In that case, the ADM masses in the $\mathrm{I}_k$ and $\mathrm{II}_k$ regions can be read off as $m$ and $-m e^{-\pi m/\alpha}$, respectively.

\section{Summary and Discussion} \label{summary}
We present a method for constructing axially symmetric static wormhole solutions supported by a phantom scalar field.
This method provides exact solutions for wormholes that connect multiple asymptotic regions.
The method we use is the gravitational soliton formalism developed by Belinski and Zakharov~\cite{Belinski1978, Belinski1979}.
A major difference from the conventional application of this formalism is that, since wormholes possess multiple asymptotic regions, it is necessary to introduce several sheets.
In this new construction, solitons $\mu_k$ corresponding to the wormholes need to be prepared, and the regularity of the connections must be checked.

We construct solitons for wormholes by referring to known spherically symmetric static solutions~\cite{Picon2002}.  
By reformulating these solutions within the soliton formalism,
we find that they can be interpreted as two-sheet wormholes described by the pair of solitons $\mu_\pm$ given in Eq.~\eqref{eq:soliton3}.  
Each soliton appears singular within a single sheet, but becomes regular once two sheets are glued together across the cuts.

Using the solitons $\mu_{\pm}$, we construct multi-sheet wormhole solutions that satisfy the regularity and asymptotically flat conditions summarized in Sec.~\ref{regularity}.
Our wormhole solutions consist of $2n$ sheets. The metric and scalar field on each sheet are given by Eqs.~\eqref{eq:fullmetricB} and \eqref{eq:soliton6phi}, and are characterized by four parameters:
the ADM mass $m$ (for the sheets $\mathrm{I}_{k}$), the scalar charge $Q$, the wormhole size $\alpha$, and the normalization constant $A$ of the spatial conformal factor.  
These parameters must satisfy the relation given in Eq.~\eqref{MSmQal} and $A = 1$, in order for the solution to meet the regularity and asymptotically flat conditions. Note that the ADM mass of the sheets $\mathrm{II}_{k}$ is given by $ - m e^{- \pi m/ \alpha}$. Thus, half of sheets have negative ADM mass.

\begin{figure}[t]
    \centering
    \includegraphics{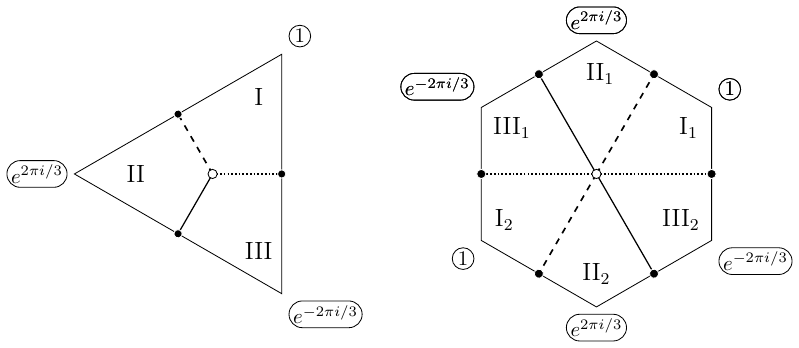}
    \caption{An example of boundary value problems. The circled values represent the phases of the complex scalar field at spatial infinity on each sheet.}
    \label{fig:oddsheets}
\end{figure}

We expect that wormhole solutions with an odd number of sheets, as well as other types of multi-sheet wormholes, can be constructed, although they are not obtained in this paper.  
The solitons for wormholes are solutions to the Laplace equation~\eqref{eq:Lap} defined on multi-sheeted spacetimes.  
As discussed in Sec.~\ref{regcut}, before imposing the condition~\eqref{eq:gauge1},  
the Laplace equation~\eqref{eq:Lap} can be regular everywhere on the multi-sheet structure.  
Since these are harmonic functions, the boundary conditions uniquely determine the solution.  
In the case of an even number of sheets, with boundary conditions where plus and minus signs appear alternately along the boundaries---as considered in this paper---the structure is similar to that of the two-sheet wormhole, and thus the same solitons $\mu_\pm$ appear.  
More generally, solitons must exist that correspond to each distinct boundary condition.  
For instance, consider the three-sheet wormhole shown in the left figure of Fig.~\ref{fig:oddsheets}.  
If we impose boundary conditions with phases $1$, $e^{2\pi i/3}$, and $e^{-2\pi i/3}$,  
the corresponding soliton exhibits $\mathbb{Z}_3$ symmetry,  
in contrast to the solitons $\mu_\pm$ introduced in this paper, which possess $\mathbb{Z}_2$ symmetry.  
If the theory involves a complex phantom scalar field, its configuration may respect $\mathbb{Z}_3$ symmetry,  
and it may be possible to construct a three-sheet wormhole or other multi-sheet wormholes based on this symmetry (see Fig.~\ref{fig:oddsheets}). 
We leave this problem for future work.

It is worth considering further generalizations, such as introducing an electromagnetic field  
or taking into account the effects of rotation.  
In the gravitational soliton formalism used to construct black hole solutions, such generalizations have already been explored.  
The introduction of an electromagnetic field would be relatively straightforward, and it is likely that the same solitons can be used.  
On the other hand, constructing rotating solutions appears to be more difficult.  
The soliton formalism developed by Belinski and Zakharov~\cite{Belinski1978,Belinski1979} was originally introduced to obtain rotating solutions.  
However, achieving regularity in these solutions tends to be accidental.  
Although a suitable soliton transformation is required, no systematic method is currently known for constructing regular rotating solutions.  
Developing such solutions remains a challenging but important direction.  
These interesting issues are left for future work.

\section*{Acknowledgment}
Y.~M. and K.~U. thank the financial supports by JST SPRING, Grant Number JPMJSP2125. 
Y.~M. and K.~U. would like to take this opportunity to thank the ``THERS Make New Standards Program for the Next Generation Researchers.'' 
K.~I. and D.~Y. are supported by Grants-in-Aid for Scientific Research from the Ministry
of Education, Culture, Sports, Science and Technology of Japan (MEXT)/ Japan Society
for the Promotion of Science (JSPS), Grant Numbers JP21H05182 (K.~I.), JP21H05189
(K.~I. and D.~Y.), JP24K07046(K.~I.), and JP20K14469 (D.~Y.).

\appendix

\section{Comparison of Solitons with Those Associated with Black Holes} \label{solitons}

Solitons $\mu_\pm$ for wormholes, as given in Eq.~\eqref{eq:soliton3}, appear to be of a different type from those for black holes.  
However, they can indeed be expressed in terms of the solitons for black holes, albeit with imaginary parameters.  
In this appendix, we present the soliton forms for wormholes in a manner similar to those for black holes.

Let us consider the Schwarzschild solution, whose line element is usually written in spherical coordinates $\qty(r, \theta, \phi)$ as
\begin{equation}
    \dd{s}^2 = -f(r)\dd{t}^2 + \frac{1}{f(r)}\dd{r}^2 + r^2\dd{\Omega}^2,
\end{equation}
where $f(r) \coloneqq 1 - 2m/r$, and the parameter $m$ corresponds to the mass of the system. This metric can be rewritten using solitons $\mu_{\pm m}$ in an axially symmetric Weyl-type form:
\begin{equation}
    \dd{s}^2 = -\frac{\mu_{-m}}{\mu_{+m}} \dd{t}^2 + \rho^2 \frac{\mu_{+m}}{\mu_{-m}} \dd{\phi}^2 + \frac{\mu_{-m}^2(\rho^2 + \mu_{+m}\mu_{-m})^2}{\mu_{+m}^2(\rho^2 + \mu_{+m}^2)(\rho^2 + \mu_{-m}^2)} \qty[\dd{\rho}^2 + \dd{z}^2],
    \label{eq:Schwarzshild in Weyl coordinates}
\end{equation}
where
\begin{equation}
    \mu_{\pm m} \coloneqq -\qty(z \mp m) + \sqrt{\rho^2 + \qty(z \mp m)^2},
\label{BHsol}
\end{equation}
and we use the coordinate transformation
\begin{equation}
    \left\{ 
    \begin{alignedat}{1}
        \rho &= \sqrt{r(r - 2m)} \sin\theta, \\
        z &= (r - m) \cos\theta.
    \end{alignedat}
    \right.
\end{equation}
Details of the construction of static solutions can be found, for instance, in Ref.~\cite{Izumi:2007qx}.

Let us return to the definition~\eqref{eq:soliton3} of the soliton $\mu_+$ for the wormhole.  
The right-hand side can be expressed as
\begin{equation}
 \Arccos w = -i \Log \left(w + i \sqrt{1 - w^2} \right),
\end{equation}
which implies that $\mu_+$ can be written as
\begin{equation}
\mu_+ = \left(w + i \sqrt{1 - w^2} \right)^{-i}.
\end{equation}
From Eq.~\eqref{def:w}, $w$ and $\sqrt{1 - w^2}$ are given by
\begin{align}
w &= \frac{1}{\rho^2} \left[ -(z + i\alpha)(z - i\alpha)
+ \sqrt{\rho^2 + (z - i\alpha)^2} \sqrt{\rho^2 + (z + i\alpha)^2} \right], \\
\sqrt{1 - w^2} &= -\frac{i}{\rho^2} \left[ (z + i\alpha) \sqrt{\rho^2 + (z - i\alpha)^2}
- (z - i\alpha) \sqrt{\rho^2 + (z + i\alpha)^2} \right].
\end{align}
The ambiguity of the square roots is fixed by requiring their real parts to be positive, i.e.,
\begin{equation}
\textrm{Re} \, \qty(\sqrt{\rho^2 + (z \pm i\alpha)^2}) > 0.
\end{equation}
Note that $\textrm{Re} \, \qty\big(\sqrt{\rho^2 + (z \pm i\alpha)^2}) = 0$ only on the cut,  
and hence the expression is smooth on each sheet.
Then, we find
\begin{equation}
\mu_{+} = \left( \frac{\mu_{+i\alpha}}{\mu_{-i\alpha}} \right)^{i},
\end{equation}
where
\begin{equation}
\mu_{\pm i\alpha} \coloneqq -\qty(z \mp i\alpha) + \sqrt{\rho^2 + \qty(z \mp i\alpha)^2}.
\end{equation}
These solitons $\mu_{\pm i\alpha}$ are exactly the same as the solitons $\mu_{\pm m}$ defined in Eq.~\eqref{BHsol} for black holes,  
except that they involve purely imaginary parameters.
The soliton $\mu_-$ is given by
\begin{equation}
\mu_{-} = e^{2\pi} \left( \frac{\mu_{+i\alpha}}{\mu_{-i\alpha}} \right)^{-i}.
\end{equation}
Therefore, wormhole solitons can be expressed in terms of the black hole solitons,  
with the distinction that the parameters are purely imaginary.

\bibliographystyle{unsrt}

\end{document}